\documentclass[journal,10pt]{IEEEtran}
\usepackage{cite}
\usepackage{amsmath,amssymb,amsfonts}
\usepackage{algorithmic}
\usepackage{graphicx}
\usepackage{textcomp}

\usepackage{microtype}
\usepackage{subfigure}
\usepackage{subcaption}
\usepackage{booktabs} 
\usepackage{float}
\usepackage{soul}
\usepackage{amsmath}
\usepackage[utf8]{inputenc}
\usepackage{svg}
\usepackage{multirow}
\usepackage[T1]{fontenc}

\def\BibTeX{{\rm B\kern-.05em{\sc i\kern-.025em b}\kern-.08em
    T\kern-.1667em\lower.7ex\hbox{E}\kern-.125emX}}
\begin{document}
\title{Leveraging Weak Supervision for Cell Localization in Digital Pathology Using Multitask Learning and Consistency Loss}
\author{Berke Levent Cesur, Ayşe Humeyra Dur Karasayar, Pinar Bulutay, Nilgun Kapucuoglu, Cisel Aydin Mericoz, Handan Eren, Omer Faruk Dilbaz, Javidan Osmanli, Burhan Soner Yetkili, Ibrahim Kulac, Can Fahrettin Koyuncu, Cigdem Gunduz-Demir
\thanks{This study was supported by Scientific and Technological Research Council of Turkey (TUBITAK) under the Grant Number 121E080. The authors thank to TUBITAK for their supports.}
\thanks{B.L. Cesur is with the Department of Computer Engineering and KUIS AI Center, Koc University, 34450 Istanbul, Turkiye (e-mail: bcesur15@ku.edu.tr).}
\thanks{A.H.D. Karasayar is with the Graduate School of Health Sciences, Koc University, 34450 Istanbul, Turkiye (e-mail: akarasayar22@ku.edu.tr).}
\thanks{P. Bulutay, N. Kapucuoglu, and C.A. Mericoz are with the Department of Pathology, Koc University, 34450 Istanbul Turkiye (e-mails: pbulutay@kuh.ku.edu.tr, nkapucuoglu@ku.edu.tr, cmericoz@kuh.ku.edu.tr).}
\thanks{H. Eren is with the Department of Pathology, Basaksehir Cam and Sakura City Hospital, 34480 Istanbul, Turkiye (e-mail: hndndlgn1993@gmail.com).
}
\thanks{O.F. Dilbaz is with the Department of Pathology, Sisli Hamidiye Etfal Health Application and Research Center, 34418 Istanbul, Turkiye (e-mail: dr.omerfarukdilbaz@gmail.com).
}
\thanks{J. Osmanli and B.S. Yetkili are with Koc University School of Medicine, Koc University, 34450 Istanbul Turkiye (e-mails: josmanli19@ku.edu.tr, byetkili19@ku.edu.tr).}
\thanks{I. Kulac is with the Department of Pathology, Graduate School of Health Sciences, Koc University Research Center for Translational Medicine, and KUIS AI Center, Koc University, 34450 Istanbul, Turkiye (e-mail: ikulac@ku.edu.tr).}
\thanks{C.F. Koyuncu is with Wallace H Coulter Department of Biomedical Engineering, Georgia Institute of Technology and Emory University, Atlanta, GA, USA (e-mail: canfkoyuncu@gmail.com).}
\thanks{C. Gunduz-Demir is with the Department of Computer Engineering, School of Medicine, and KUIS AI Center, Koc University, 34450 Istanbul, Turkey (e-mail: cgunduz@ku.edu.tr).}}

\maketitle







\begin{abstract}
Cell detection and segmentation are integral parts of automated systems in digital pathology. Encoder-decoder networks have emerged as a promising solution for these tasks. However, training of these networks has typically required full boundary annotations of cells, which are labor-intensive and difficult to obtain on a large scale. However, in many applications, such as cell counting, weaker forms of annotations--such as point annotations or approximate cell counts--can provide sufficient supervision for training. This study proposes a new mixed-supervision approach for training multitask networks in digital pathology by incorporating cell counts derived from the eyeballing process--a quick visual estimation method commonly used by pathologists. This study has two main contributions: (1) It proposes a mixed-supervision strategy for digital pathology that utilizes cell counts obtained by eyeballing as an auxiliary supervisory signal to train a multitask network for the first time. (2) This multitask network is designed to concurrently learn the tasks of cell counting and cell localization, and this study introduces a consistency loss that regularizes training by penalizing inconsistencies between the predictions of these two tasks. Our experiments on two datasets of hematoxylin-eosin stained tissue images demonstrate that the proposed approach effectively utilizes the weakest form of annotation, improving performance when stronger annotations are limited. These results highlight the potential of integrating eyeballing-derived ground truths into the network training, reducing the need for resource-intensive annotations.
\end{abstract}


\section{Introduction}
In digital pathology, the development of automated decision support systems is essential for enabling high-throughput analysis and reducing intra- and inter-observer variability. A core component of these systems is automated cell segmentation or detection, which serves various purposes. For example, segmentation/detection outputs can be used directly to count cells in tissue samples or as inputs for downstream tasks, such as constructing cell graphs and applying graph neural network classifiers. The most recent approach for localizing cells in tissue images is to develop dense prediction networks~\cite{ronneberger2015Unet,chen2021transunet,fujita2020maskrcnn}. When designed for segmentation, these networks typically require strong boundary annotations of the cells in the training data. The demand for such data increases significantly when training large-scale segmentation models~\cite{graham2019Hovernet,greenwald2022wholecell}. However, delineating cell boundaries in tissue images is one of the most labor-intensive tasks in digital pathology, making it highly challenging to acquire boundary annotations for a large number of cells. While segmentation networks and detailed boundary annotations are necessary for applications that extract histomorphological features from individual cells, many other applications, such as cell counting, only require the approximate localization of the cells. Thus, weaker forms of annotation may be sufficient to train networks for such applications. Moreover, these weak annotations can be used together with strong annotations to better train segmentation networks. 

In cell microscopy imaging, researchers have explored the use of various weak annotations to train segmentation and detection networks in weakly- or mixed-supervised learning approaches. Common forms of weak annotations include scribbles~\cite{oh2022scribble}, bounding boxes~\cite{aldughayfiq2023yolov5}, and, most prominently, point annotations~\cite{khalid2023pace,miao2021quick}, which involve placing a single point within each cell. Numerous studies developed alternative training techniques to utilize point annotations. For example, the studies in~\cite{yu2023point} and~\cite{zhao2021weakly} employed self- and co-training techniques to generate cell segmentation masks from single point annotations. The latter study~\cite{zhao2021weakly} also integrated a human-in-the-loop mechanism, enabling additional annotations to finetune the trained network. Other studies~\cite{pan2018cell,xie2018efficient} localized cells in an image by training regression networks to learn density maps of the single point annotations. Unlike dense prediction networks trained with strong boundary annotations, these regression models required preprocessing of point annotations using techniques such as Gaussian kernels~\cite{qu2020weakly} and~repel coding~\cite{chamanzar2020weakly}. The regression-based models were particularly advantageous in cases involving overlapping cells in densely populated cultures with background clutter. 


Beyond cell localization, another important practical application of digital pathology is to estimate the number of cells in a tissue region, which can provide as a significant indicator, e.g., clinicians may infer a patient's cancer stage or type by monitoring variations in cell counts~\cite{chow2020countingmitoses,frei2023pathologist}. Cell counts can be derived from the outputs of both segmentation networks and regression-based models. For example, the number of connected components in a map estimated by a segmentation network or the sum of pixel values in a density map estimated by a regression network can be used to infer cell counts. Alternatively, cell counts can be directly used as supervisory signals to train models. In the literature, cell count data has been used either as the sole ground truth to train convolutional neural networks~\cite{xue2016cellcounting,lavitt2021deep} or as an auxiliary supervisor signal alongside the cell segmentation task in multitask networks~\cite{hagos2019concordenet,huang2022msrcn}. Multitask networks were designed with separate branches for each task, enabling joint learning from shared representations. Because of this joint learning to optimize shared parameters, multitask learning not only enhances the model's generalization ability~\cite{graham2023one,alom2020mitosisnet} but also has great potential in data utilization, especially when annotated data is scarce for one task but abundant for another, attribute that makes it an ideal approach for mixed supervised settings as in the case of our proposed methodology.

Previous studies that incorporated cell counts into their model training assumed that these counts were precise. However, determining the exact number of cells requires counting them one by one, which is nearly as labor-intensive as obtaining point annotations. On the other hand, in typical clinical settings, pathologists often estimate cell counts using a method known as \textit{eyeballing}~\cite{talat2023ki67}, which involves a preliminary visual assessment of a tissue region and an approximate estimate of the number of cells in that region by sight. This process takes significantly less time than obtaining boundary or point annotations, and the cell count obtained at the end of this process can be considered as the simplest/weakest form of annotation for cell localization and counting tasks. Despite that, \textit{the use of eyeballing-derived ground truths for supervising cell count predictions remains largely unexplored, both as a standalone cell counting task and as an auxiliary task for cell localization.}

To address this gap, this paper proposes a new mixed-supervised training scheme for a multitask network to exploit ground truths obtained by the eyeballing process, reflecting a more realistic setting for acquiring cell count annotations in digital pathology. In the proposed multitask network, each branch is dedicated to a different task, which is trained with different supervisory signals determined by the level of annotations available for a training image. In our setting, ground truth cell counts are obtained by eyeballing for all training images, but point annotations are only available for a smaller subset. Since the obtained cell counts are only rough numbers and may be erroneous due to eyeballing, this paper also introduces a new loss function with a consistency term that penalizes inconsistencies between the cell counts inferred from the predictions of the two branches, in particular, to regularize learning for the cell counting task. The main contributions of this paper are twofold:
\begin{itemize}
\item This is the first proposal of a mixed-supervision approach for digital pathology that utilizes cell counts obtained by eyeballing in a multitask network to simultaneously learn the tasks of cell localization and cell counting. 
\item A new consistency term is introduced as a regularization technique to enhance task coherence in a multitask network. This term is defined to penalize differences between the predictions of the two tasks, specifically the difference between the number of cells predicted by the cell counting task and the number of cell objects (connected components) in the segmentation map predicted by the cell localization task. Our proposal is different than the previous studies that defined this term between the outputs of multiple cell localization networks, which were designed to predict the same task. As cell localization and cell counting are relevant but more diverse tasks, they are more complementary and the consistency term defined between the predictions of these tasks is expected to be more regulative. 
\end{itemize}

Testing the proposed methodology in two datasets of tissue images stained with hematoxylin-eosin (HE), our experiments showed that the utilization of cell counts obtained by eyeballing in the context of multitask learning improved the performance of the model under the scarcity of stronger annotations.

\section{Related Work}
This study proposes a mixed-supervised learning scheme that utilizes point annotations and cell counts as supervisory signals to train a multitask network. It also introduces a loss with a consistency term to enhance task coherence. In the following, we categorize related work based on these aspects. 

\textbf{\underline{Cell localization with point annotations}:} Point annotations have been utilized in two primary ways to train deep learning models. The first approach involves training a regression network to estimate density maps generated from the point annotations, enabling the model to learn cell locations. The second approach leverages these points to generate cell segmentation masks, providing a more detailed understanding of cellular structures. The first group employed a common approach to generate density maps, utilizing Gaussians~\cite{qu2020weakly,nishimura2021weakly} and proximity functions~\cite{xie2018efficient,pan2018cell}. In this method, the points were treated as peaks, with densities decreasing as one moved away from these points. To learn these density maps from the original images, a regression-based density prediction network was trained, enabling the model to accurately predict the density maps from the input data. In~\cite{qu2020weakly}, the predicted Gaussian masks were enhanced by obtaining Voronoi diagrams from the detected regions and combining them with pixel-level k-means clustering. Similarly, in~\cite{chamanzar2020weakly}, the repel encoding was used together with Voronoi diagrams and pixel clustering to improve cell localization.

The second group of studies focused on generating segmentation maps from point annotations, often utilizing self- and co-training networks~\cite{yu2023point,zhao2021weakly,lin2023nuclei}. For instance, a study proposed a self-learning approach that shared knowledge between a principal and a collaborator model~\cite{yu2023point}. The primary model was trained for object detection using point annotations, while the collaborator model was trained for semantic segmentation under the guidance of the primary model. Another study~\cite{lin2023nuclei} employed a self-supervised learning approach, where point annotations were used as starting points to train a segmentation model. The initial step involved converting the point annotations into rough pixel-level annotations based on Voronoi diagrams and k-means clustering, followed by a progressive approach to refine the labels. In another work~\cite{Yoo2019PseudoEdgeNetNS}, a nucleus segmentation method was developed solely relying on point annotations. To address undersegmentation, a secondary task was defined and learned with an additional small network. In ~\cite{khalid2022point2mask}, point annotations were combined with cells' bounding boxes to generate a segmentation mask. Although this approach required additional bounding box annotations, it was argued that this level of annotation was still less labor-intensive than obtaining full boundary annotations. Our methodology differs significantly from these existing approaches. We propose training a multitask network with different levels of supervision to predict cell locations from point annotations. Furthermore, we incorporate cell counts obtained by eyeballing into the training process, providing a unique and innovative approach to cell location prediction.

\textbf{\underline{Deep learning models for cell counting}:} One group of studies calculated cell counts by analyzing the outputs predicted by cell segmentation or detection networks. In the case of cell segmentation, the count was determined by simply counting the number of objects present on the segmentation map~\cite{dave2023mimo,melouk2023autoencoders,chen2021cell-localization,morelli2021cResUnet}. Similarly, for detection networks, mostly based on YOLO models, cell counts were obtained by counting the bounding boxes generated around the detected cells~\cite{tessema2021quantitative,aldughayfiq2023yolov5,bai2023blood}.

Another group of studies tackled the cell count prediction problem by framing it as a regression task. They utilized point annotations to generate density maps, which were then used to train a regression network. The trained network predicted the density maps, and cell counts were calculated by integrating the pixel predictions~\cite{xie2016microscopy,he2021deeply,zhu2021realtime,guo2019saunet}.

Alternatively, some studies predicted cell counts by feeding the predicted density maps~\cite{liu2019automated,liu2019novel,hagos2019concordenet} to a second regression network with convolutional and dense layers~\cite{vununu2018deep}. These networks were either single-task networks trained using cell count information as the sole supervisory signal~\cite{xue2016cellcounting, lavitt2021deep} or multitask networks that concurrently learned cell count prediction from a shared encoder together with additional tasks of cell segmentation~\cite{huang2022msrcn} or spatial super-resolution~\cite{deng2023ssrnet}. Our proposed approach is similar to these multitask networks in that it includes multiple branches for the tasks of cell localization and cell count prediction. However, our approach introduces a consistency term in its loss function to ensure coherence in the collaborative learning of its tasks. This allows our model to use cell counts obtained from eyeballing for supervision, rather than requiring exact counts.

\textbf{\underline{Forcing consistency through a loss function}:} It was used as a regularization technique to enforce consistent predictions from multiple models trained on the same data. It helped models make stable predictions, which usually led to better generalization and performance especially when annotated data was scarce~\cite{peng2021medical}. The study in~\cite{wu2022mutualconsistency} proposed a multi-task network for radiologic image segmentation, defining a mutual consistency constraint between the probability map of one task and the pseudo-labels of the others to reduce uncertainty in semi-supervised image segmentation. 

The consistency term can also be defined between the outputs of different tasks. For example, a multitask network was proposed for endoscopic image classification and segmentation with a cross fusion module, calculating a consistency loss between the features of the classification task to reduce the misalignment between classification and segmentation results~\cite{kong2021multitask}. Consistency loss functions were also used to train cell segmentation networks. In~\cite{zhao2021weakly}, a cell segmentation network was trained with weak supervision using self- and co-training schemes, with a consistency loss to ensure consensus between a self-trained and two co-trained networks. In~\cite{yu2023point}, cell segmentation was achieved using a principal and a collaborator model, both trained through a self-learning strategy that involved knowledge sharing between the models. This knowledge sharing was achieved by defining a consistency loss between the predictions of the two models. Unlike the previous studies that defined their consistency loss on the results of two segmentation tasks, our proposed method defines it between the prediction of the cell counting task and the number of components in the segmentation map predicted by the cell localization task.

\begin{figure*}[t]
\centering
\includegraphics[width=17cm]{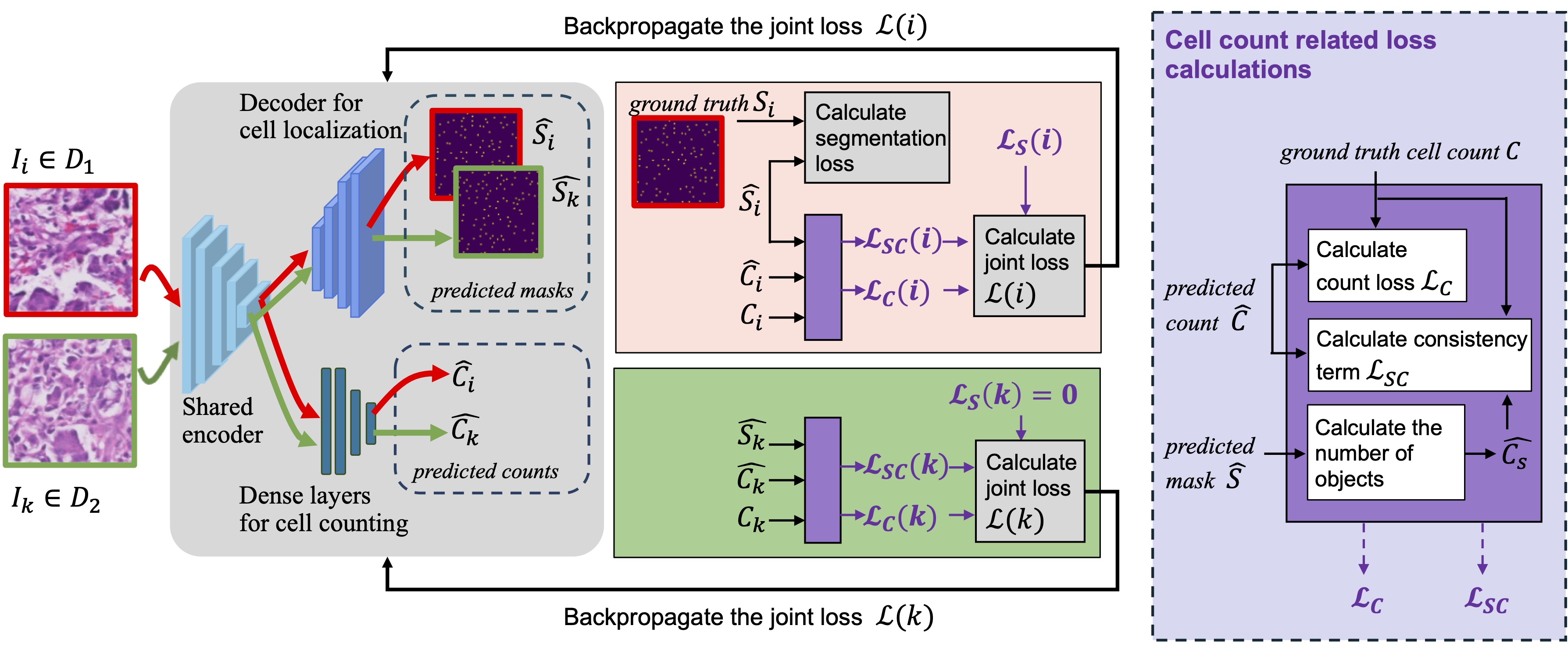}
\caption{Overview of the proposed mixed-supervised strategy to train the multitask network with different levels of supervision. In particular, for a training image $I_i \in D_1$, the ground truth mask $S_i$ generated from point annotations and the cell count $C_i$ obtained by counting the annotated points are available, and all loss components are calculated between the ground truths and the predicted values $\hat{S_i}$ and $\hat{C_i}$ (red boxes and arrows). For a training image $I_k \in D_2$, only the cell count $C_k$ obtained by eyeballing is available, and only cell count related loss components are calculated (green boxes and arrows). Note that for $I_k$, one can also calculate the consistency term ${\cal L}_{SC}(k)$ since this calculation uses the ground truth count $C_k$, its predicted value $\widehat{C_k}$, and the number $\widehat{C_{s_k}}$ of connected components in the predicted mask $\widehat{S_k}$, but not the ground truth mask $S_k$, which is not available for $I_k$. This calculation is depicted in the purple box on the right. After calculating the joint loss, relevant network weights are updated through backpropagation.}
\label{fig:training_pipeline}
\end{figure*}

\section{Methodology}
This paper proposes a mixed supervised training strategy for a multitask network to concurrently learn the tasks of cell localization and cell counting. This network has one shared encoder and two separate branches, one decoder for cell localization and one fully connected layer for cell counting. The proposed strategy is to train the multitask network when different levels of supervision are available for different instances in a training dataset. It assumes that the training dataset $D = {(I_n,Y_n)}$, with $I_n$ being the input image and $Y_n$ corresponding to the ground truth annotations available for this image, consists of two subsets $D = D_1 \cup D_2$. The first subset, $D_1 = \{I_i \,|\, Y_i \in (S \cup C)\}$, contains training images $I_i$ with two types of ground truth annotations: binary segmentation masks $S_i\in \{0,1\}$ obtained by dilating point annotations, and cell counts $C_i \in \mathbb{Z}^+$ obtained by counting the annotated points. The second subset, $D_2 = \{I_k \,|\, Y_k \in C\}$, consists of training images $I_k$ for which only the cell counts $C_k \in \mathbb{Z}^+$ are available, estimated by expert pathologists using the conventional eyeballing method. These images do not have precise cell locations or segmentation masks, but rather a rough estimate of the number of cells present in each image.

Our training strategy employs a two-stage approach to update the network weights. When $I_n \in D_1$, all network weights are updated simultaneously. However, when $I_n \in D_2$, only the weights of the shared encoder and the decoder associated with the cell counting task are updated. This selective weight update is achieved by backpropagating the joint loss, which includes the proposed consistency term, through the network.

The overview of the proposed methodology is illustrated in Fig.~\ref{fig:training_pipeline} and the details are provided in the following subsections.

\subsection{Multitask Network Architecture}
\begin{figure*}[t]
    \centering
    \includegraphics[width=17cm]{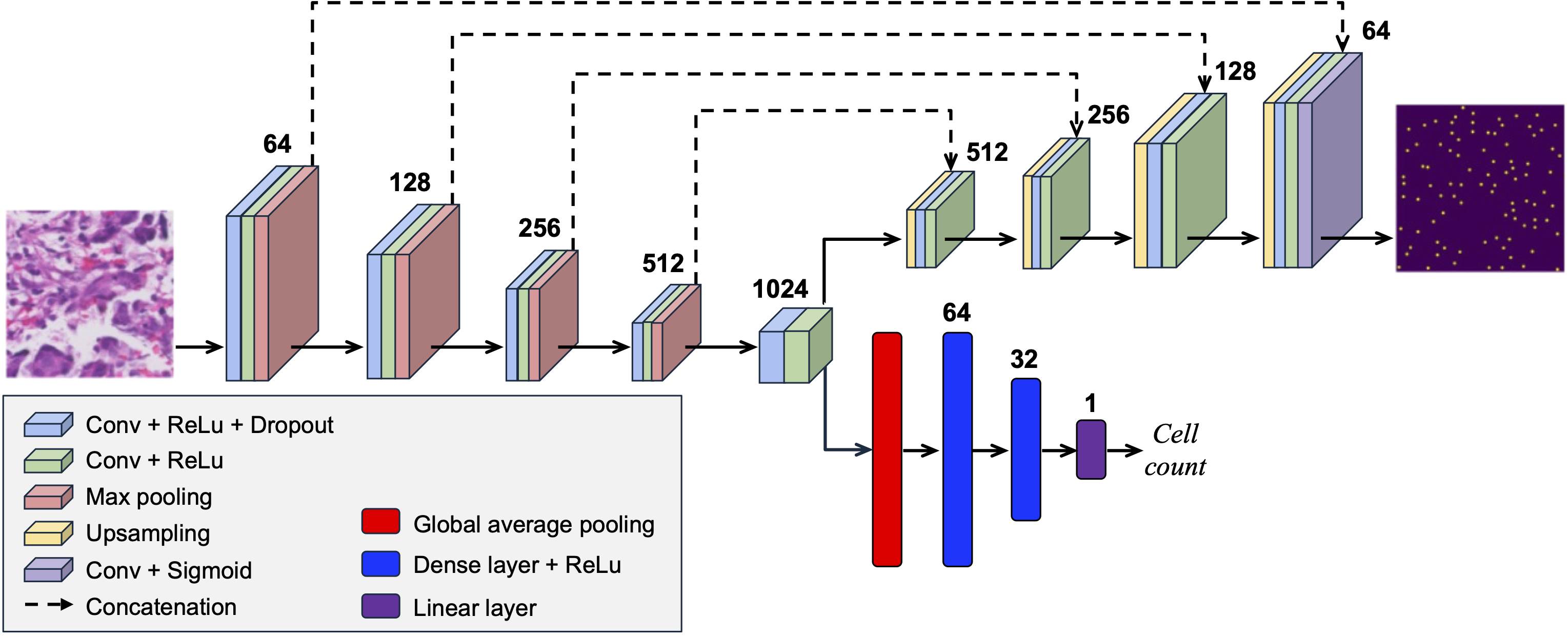}
    \caption{Architecture of the proposed multitask network. The number of feature maps used in each block is indicated on its top.}
    \label{fig:model-architecture}
\end{figure*}

The multitask network consists of a shared encoder accompanied by a decoder for cell localization and fully connected layers for cell count prediction (Fig.~\ref{fig:model-architecture}). Its encoder consists of four blocks, each containing two convolutional and one max pooling layer. The convolutions use a $3\times3$ kernel and a rectified linear unit (ReLu) as the activation function, while the max pooling employs a $2\times2$
kernel to reduce the spatial dimension. To prevent overfitting, a dropout layer with a rate of 0.2 is added between two consecutive convolutional layers. The bottleneck contains two convolutions and a dropout layer between them. The decoder also consists of four blocks, where in each block, upsampling with a $2\times2$ kernel is applied to recover the original input size, and its output is concatenated with the features of the corresponding encoder block before being fed into a convolution with a $3\times3$ kernel. At the end, the sigmoid function is used to obtain binary pixel labels. In the branch of cell count prediction, global average pooling is first applied to the features obtained from the shared encoder, followed by two dense layers with 32 and 16 hidden units, respectively, both using a ReLu as the activation function. At the end, the linear function is used to obtain cell counts, as it corresponds to a regression task. 

\subsection{Settings for Separate Branches}

The first branch is for dense prediction to locate cells on an image. It is supervised by ground truth masks that contain rough locations of the cells. These masks are generated by dilating point annotations provided by pathologists. For any given ground truth mask $S_n$, the pixel value is 1 if a pixel belongs to the cell area after dilation, and 0 otherwise. For an image $I_n \in D_1$, the loss ${\cal L}_S(n)$ of this branch is the binary cross entropy, which is defined between the pixel values in $S_n$ and those in the segmentation map $\widehat{S_n}$ predicted by the network. It is worth noting that this loss can be calculated for training images $I_n \in D_n$ with point annotations, and the weights for the decoder of this branch are updated. For the other training images $I_n \in D_2$, the loss of this branch ${\cal L}_S(n)$ is set to 0, since there are no point annotations to generate the ground truth masks, and thus, there is no update of the decoder's weights for these training images. 

The second branch corresponds to a regression problem for cell count prediction. It is supervised by the ground truth count $C_n$ obtained by counting the annotated points for the images in the first set, $I_n \in D_1$, and by eyeballing for those in the second set, $I_n \in D_2$. For an image $I_n$, the loss ${\cal L}_C(n)$ of this branch is defined as
\begin{equation}
{\cal L}_C(n) = \frac{|C_n - \widehat{C_n}|}{C_n}
\end{equation}
where $\widehat{C_n}$ is the cell count predicted by the network. It is an absolute error between the ground truth and the predicted cell counts, normalized by the ground truth. This normalization ensures that the loss is comparable for all images, regardless of the number of cells they contain. This loss is defined for all training images in $D = D_1 \cup D_2$ because it does not require knowing the point annotations. Hence, the weights of the dense layers of this branch are updated for all training images. 

\subsection{Mixed-Supervised Network Training with Joint Loss}

The two tasks of cell localization and cell count prediction concurrently learned by optimizing the joint loss function ${\cal L}(n)$ with the proposed consistency term. It is defined as
\begin{equation}
{\cal L}(n) = \left\{ 
\begin{aligned} 
&  {\cal L}_S(n) &+ \alpha~{\cal L}_C(n) + \beta~{\cal L}_T(n),
& \text{~~~if~}I_n \in D_1 \\ 
&  0&+\alpha~{\cal L}_C(n) + \beta~{\cal L}_T(n),
& \text{~~~if~}I_n \in D_2 \\ 
\end{aligned} \right.
\end{equation}
where ${\cal L}_T(n)$ is the consistency term for the image $I_n$, and $\alpha$ and $\beta$ are the relative contributions of the cell count loss and the consistency term to the joint loss function, respectively. The consistency term for the image $I_n$ is defined as
\begin{equation}
{\cal L}_T(n) = \frac{|\widehat{C_n} - \widehat{C_{s_n}}|}{C_n}
\end{equation}
where $C_n$ is the ground truth, $\widehat{C_n}$ is the cell count predicted by the cell counting (second) branch of the network, and $\widehat{C_{s_n}}$ is the number of cell objects (connected components) in the mask $\widehat{S_n}$ predicted by the cell localization (first) branch of the network. This term can also be calculated for all training images in $D = D_1 \cup D_2$ because it does not use the ground truth mask $S_n$, but the predicted mask $\widehat{S_n}$. As the dense layers of the cell count prediction task are optimized using the weakest form of supervision, which may contain error due to the nature of eyeballing, this consistency term is used to update the weights of the dense layers of this task to regularize its predictions. Additionally, since the cell localization branch may produce noisy masks in early epochs, the consistency term is used after a warm-up period, it is not used in the first 25 epochs. The weights of the shared encoder are updated by backpropagating all loss terms available for a given training image. 

All networks are constructed and trained in the Python programming language using TensorFlow. The Adam optimizer is used to train the networks. The learning rate and exponential decay parameters are set to 0.0001 and 0.9, respectively, and the batch size is 1. Training continues for a maximum of 200 epochs, and the network weights that give the best loss value in the validation dataset are selected.

\section{Experiments}

\subsection{Datasets}
We conducted our experiments on two datasets of HE stained tissue images. The first is the in-house serous carcinoma dataset, comprising 109 images with a resolution of $1536 \times 1536$ pixels. These images were cropped from multiple whole slide images of various subjects. The data collection was conducted in accordance with the tenets of the Declaration of Helsinki and approved by Koc University Institutional Review Board (protocol number: 2021.336.IRB2.066). Point annotations of these images were performed by a team of pathologists. After months of obtaining these point annotations, a pathologist was shown each image and asked to estimate the cell count by eyeballing. Although both annotations were available for all images, when training a network, we used one annotation or the other, depending on how we partitioned the training dataset into $D_1$ and $D_2$. When $I_n \in D_1$, we generated a segmentation map by dilating the point annotations with a disk structuring element. The disk radius was selected as 3 considering the average cell size in the images of this dataset. Then, we used the generated segmentation map as the ground truth mask $S_n$ and the number of the annotated points as the ground truth count $C_n$. When $I_n \in D_2$, we used the cell count obtained by eyeballing as the only ground truth $C_n$. To understand the effectiveness of the proposed training strategy against data scarcity, we performed our experiments with different partition percentages, each time putting $p$ percent of the training data into the first dataset $D_1$, and the rest into the weakly annotated dataset $D_2$. For the in-house serous carcinoma dataset, we used three-fold cross validation. In each trial, we used all images in two folds for training; we used approximately 80 percent of the images to learn the network weights and 20 percent as validation data to select the best network weights. We used the other fold as a test set to evaluate the model performance. We calculated the performance metrics on the test images using their cell counts obtained by counting the point annotations, rather than those obtained by eyeballing, since this would gave more reliable results.

The second dataset, MoNuSeg, is a publicly available collection of HE-stained tissue images from various organs and patients~\cite{kumar2017dataset}. The dataset originally consisted of 37 training and 14 test images. We maintained the original training and test splits, using 20 percent of the training images as validation data. To ensure compatibility with our multitask network's architecture, we cropped four non-overlapping patches of $496\times496$ pixels from each original image, which had a resolution of $1000\times1000$. We conducted multiple experiments by dividing the training data into two sets, $D_1$ and $D_2$, with varying percentages of data, $p$, in each set.
The MoNuSeg dataset originally included cell boundary annotations for all images. However, to simulate weaker supervision with point annotations and eyeballing, we did not use these boundary annotations. Instead, we calculated the centroids of the cells and used them as point annotations. The ground truth masks were then generated by dilating these point annotations. Additionally, a team of pathologists provided cell counts for each image by eyeballing.

\begin{table}[t!]
\centering
\caption{The numbers of images and cells in the training, validation, and test sets. Since we used three-fold cross validation for the serous carcinoma dataset, we provided these numbers for each trial separately.}
\begin{tabular}{|l|c|c|c|c|c|c|}
\hline
\multirow{2}{*}{Dataset} & \multicolumn{3}{c|}{Number of images} & \multicolumn{3}{c|}{Number of cells} \\ \cline{2-7}
& Tr & Val & Ts & Tr & Val & Ts \\ \hline
Serous (trial 1)  & 59  & 14 & 36 & 65711 & 15895 & 43025 \\ \hline
Serous (trial 2)  & 58  & 14 & 37 & 68251 & 17497 & 38883 \\ \hline
Serous (trial 3)  & 58  & 15 & 36 & 66509 & 15399 & 42723 \\ \hline
MoNuSeg           & 120 & 28 & 56 & 16794 &  7180 &  6619 \\ \hline
\end{tabular}
\label{table:folds_stats}
\end{table}

\begin{figure}[t]
\centering
\begin{tabular}{c@{~~}c}
\includegraphics[width=4cm]{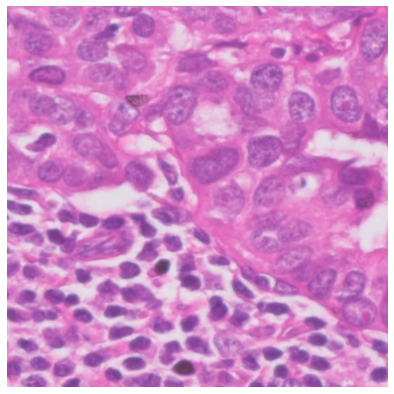} & 
\includegraphics[width=4cm]{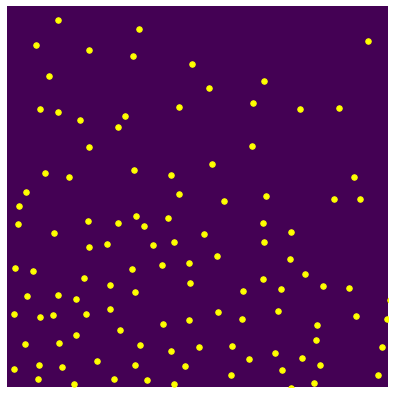} 
%
\end{tabular}
\caption{An example patch, with $384\times 384$ pixel resolution, cropped from an original image, with $1536\times 1536$ resolution, of the serous carcinoma dataset, and its point annotations. As shown here, obtaining point or boundary annotations is challenging when there are many cells to annotate. It is worth noting that this patch is 1/16th of the original image.}
\label{fig:dataset}
\end{figure}

The numbers of images and cells in the training, validation, and test sets for each dataset are provided in Table~\ref{table:folds_stats}. It is worth noting that we performed three-fold cross-validation on the in-house dataset, while using the original training and test splits for the MoNuSeg dataset. The serous carcinoma dataset presents a unique challenge due to its high cell density, making point or boundary annotations particularly difficult. In such cases, the weakest form of annotation, i.e., cell counts by eyeballing, becomes a more efficient option. To illustrate this challenge, Fig.~\ref{fig:dataset} shows a small patch cropped from an original image of the serous carcinoma dataset. This patch represents only 1/16th of the original image, which has a resolution of $384 \times 384$ and $1536 \times 1536$ pixels, respectively.

\subsection{Evaluations}

We evaluated the performance of our model on cell localization and counting tasks using a test set/fold. For each test image, we calculated various performance metrics and averaged them over the entire test set/fold. For cell localization, we identified true positive (TP) cells in each predicted map and calculated the following metrics: object-level precision, object-level recall, and F1-score. To identify TP cells, we matched each point annotation with predicted objects within a distance threshold of 10 pixels. A point annotation $p_i$ was considered a TP if it had exactly one matching object $q_j$ that was not matched with any other point annotation. The distance threshold of 10 pixels was empirically selected considering the average cell size and the image resolution. For the MoNuSeg dataset, we removed regions smaller than 35 pixels from the predicted maps to eliminate noisy predictions. This postprocessing step was applied to both our model's predictions and those of comparison algorithms. In contrast, the serous carcinoma dataset did not require any postprocessing.

For cell count prediction, we used the relative difference as the performance metric. For each test image $I_n$, it is defined as $|C_n - \widehat{C_n}| / C_n$, where $C_n$ is the ground truth and $\widehat{C_n}$ is the predicted cell count. In calculating this metric, we used the number of annotated points as the ground truth $C_n$, rather than relying on the count obtained by eyeballing. We calculated and reported the relative difference metric by considering the predictions of each branch separately. First, we calculated it on the map predicted by the cell localization branch, where $\widehat{C_n}$ was taken as the number of the objects (connected components) in the predicted map. We then used the direct estimation from the cell count prediction as $\widehat{C_n}$. We will refer to these metrics as $\text{RD}_\text{Count}$ and $\text{RD}_\text{Loc}$, respectively. 

\begin{table*}[t]
\centering
\caption{For the serous carcinoma dataset, the test fold metrics obtained by the proposed \textit{MixedSupervision} model, its single-task counterparts, and when the consistency term is not used in the joint loss function. Results are the averages of 9 runs from 3 folds and their standard deviations.}
\begin{tabular}{|c|c|c|c|c|c|c|c|c|c|}
\hline
\multirow{3}{*}{$p$\%} & \multicolumn{3}{c|}{} & \multicolumn{2}{c|}{Single-task} & Single-task & \multicolumn{3}{c|}{}\\ 
 & \multicolumn{3}{c|}{\textit{MixedSupervision}} & \multicolumn{2}{c|}{cell localization} & cell counting & \multicolumn{3}{c|}{\textit{MixedSupervision} -- w/o consistency}\\ \cline{2-10}
 & F1-score & $\text{RD}_\text{Loc}$ & $\text{RD}_\text{Count}$ & F1-score & $\text{RD}_\text{Loc}$ & $\text{RD}_\text{Count}$ & F1-score & $\text{RD}_\text{Loc}$ & $\text{RD}_\text{Count}$\\ \hline
 100 & 84.45 $\pm$ 1.84 & 0.14 $\pm$ 0.04 & 0.19 $\pm$ 0.05 & 83.34 $\pm$ 1.66 & 0.13 $\pm$ 0.04 & 0.24 $\pm$ 0.02 & 83.69 $\pm$ 1.82 & 0.14 $\pm$ 0.03 & 0.24 $\pm$ 0.11\\
 75 & 83.87 $\pm$ 1.80 & 0.13 $\pm$ 0.03 & 0.20 $\pm$ 0.06 & 82.15 $\pm$ 1.40 & 0.14 $\pm$ 0.03 & 0.24 $\pm$ 0.03  & 82.69 $\pm$ 1.90 & 0.15 $\pm$ 0.04 & 0.35 $\pm$ 0.20\\
 50 & 82.77 $\pm$ 1.82 & 0.14 $\pm$ 0.02 & 0.22 $\pm$ 0.02 & 82.02 $\pm$ 1.83 & 0.14 $\pm$ 0.02 & 0.26 $\pm$ 0.02 & 82.16 $\pm$ 2.16 & 0.15 $\pm$ 0.03 & 0.41 $\pm$ 0.21\\
 25 & 81.45 $\pm$ 1.95 & 0.15 $\pm$ 0.03 & 0.28 $\pm$ 0.05 & 80.08 $\pm$ 1.63 & 0.16 $\pm$ 0.03 & 0.29 $\pm$ 0.02 & 80.99 $\pm$ 1.39 & 0.16 $\pm$ 0.03 & 0.44 $\pm$ 0.12\\\hline
\end{tabular}
\label{table:serous-table}
\end{table*}

\begin{table*}[t]
\centering
\caption{For the MoNuSeg dataset, the test set metrics obtained by the proposed \textit{MixedSupervision} model, its single-task counterparts, and when the consistency term is not used in the joint loss function. Results are the averages of 3 runs and their standard deviations.}
%
\begin{tabular}{|c|c|c|c|c|c|c|c|c|c|}
\hline
\multirow{3}{*}{$p$\%} & \multicolumn{3}{c|}{} & \multicolumn{2}{c|}{Single-task} & Single-task & \multicolumn{3}{c|}{}\\ 
 & \multicolumn{3}{c|}{\textit{MixedSupervision}} & \multicolumn{2}{c|}{cell localization} & cell counting & \multicolumn{3}{c|}{\textit{MixedSupervision} -- w/o consistency}\\ \cline{2-10}
 & F1-score & $\text{RD}_\text{Loc}$ & $\text{RD}_\text{Count}$ & F1-score & $\text{RD}_\text{Loc}$ & $\text{RD}_\text{Count}$ & F1-score & $\text{RD}_\text{Loc}$ & $\text{RD}_\text{Count}$\\ \hline
 100 & 80.86 $\pm$ 0.44 & 0.07 $\pm$ 0.01 & 0.14 $\pm$ 0.02 & 78.57 $\pm$ 0.53 & 0.08 $\pm$ 0.01 & 0.28 $\pm$ 0.02 & 79.57 $\pm$ 1.07 & 0.07 $\pm$ 0.01 & 0.22 $\pm$ 0.12 \\
 75 & 79.10 $\pm$ 0.55 & 0.07 $\pm$ 0.03 & 0.15 $\pm$ 0.03 & 77.60 $\pm$ 1.07 & 0.09 $\pm$ 0.02 & 0.31 $\pm$ 0.02 & 77.86 $\pm$ 2.23 & 0.08 $\pm$ 0.01 & 0.22 $\pm$ 0.07 \\
 50 & 77.68 $\pm$ 0.28 & 0.09 $\pm$ 0.03 & 0.19 $\pm$ 0.03 & 75.17 $\pm$ 1.85 & 0.12 $\pm$ 0.02 & 0.32 $\pm$ 0.04 & 75.14 $\pm$ 1.56 & 0.10 $\pm$ 0.02 & 0.33 $\pm$ 0.12 \\ 
 25 & 76.41 $\pm$ 0.11 & 0.09 $\pm$ 0.03 & 0.18 $\pm$ 0.01 & 73.34 $\pm$ 1.16 & 0.11 $\pm$ 0.04 & 0.47 $\pm$ 0.02 & 73.93 $\pm$ 1.04 & 0.10 $\pm$ 0.02 & 0.50 $\pm$ 0.06 \\\hline
\end{tabular}
\label{table:monuseg-table}
\end{table*}

\subsection{Comparison Algorithms}

We compared our proposed training approach with two state-of-the-art models: ConCORDe-Net~\cite{hagos2019concordenet} and SSR-Net~\cite{deng2023ssrnet}. Both models are designed for the dual tasks of cell detection and counting.

ConCORDe-Net has two modules: one for cell detection and another for cell counting. Unlike our approach, it first detects cells using an encoder-decoder network and then uses the generated cell detection mask as input to the cell counting module. We implemented the cell detection module using the same encoder-decoder network architecture and loss function as ConCORDe-Net to ensure a fair comparison. For the cell count prediction module, we used the loss function suggested by~\cite{hagos2019concordenet}, which calculates a similarity measure by inverting the mean absolute error between the ground truth and predicted cell counts over the batch, and then subtracting it from 1 to obtain a final loss value between 0 and 1. Note that this loss function does not include any consistency term.

SSR-Net is a multitask learning framework that learns cell count prediction within a multitask network for cell segmentation and cell count prediction. It generates a cell segmentation/detection map by reconstructing the input image with a spatial-based super-resolution module, and regresses the cell count from the features extracted through the upsampling module of this reconstruction. SSR-Net is promoted as a non-point-based counting method. In our experiments, instead of using point annotations to generate cell segmentation masks, we trained their model to reconstruct the input images, providing both cell segmentation maps and cell count predictions as outputs.

\section{Results and Discussion}

We trained our proposed approach (\textit{MixedSupervision}) on a training set $D = D_1 \cup D_2$ with mixed supervisory signals. Specifically, $D_1$ contains point annotations, with the number of annotated points serving as the cell counts, and $D_2$ contains only the cell counts obtained by eyeballing. In our experiments, we used $p$ percent of the training images in $D_1$ and the rest in $D_2$, where $p$ determines the percentage of data requiring labor-intensive annotation. The primary goal of this work is to design a model that reduces the annotation effort required for learning cell localization and counting tasks. To investigate this, we first analyzed the effects of different $p$ percentages on training performance. Tables~\ref{table:serous-table} and~\ref{table:monuseg-table} show the results for the serous carcinoma and MoNuSeg datasets, respectively. Since network weight initialization affects the final model, we repeated each experiment three times for our model, as well as for the comparison algorithms and all ablation studies. The values in Table~\ref{table:serous-table} represent the average test fold results from nine runs (three folds, with three runs per fold) and Table~\ref{table:monuseg-table} reports the average test set results from three runs (as there was a single test set for the MoNuSeg dataset).

These tables also report comparative results for the single-task counterparts. We trained two single-task networks. The first one learned cell localization from the masks generated using the point annotations available in $D_1$, which contained only $p$ percent of the entire training set. After training, we calculated the object-level F1-score on the prediction masks of the test images, and also reported $\text{RD}_\text{Loc}$, which was calculated as the relative difference between the ground truth and the number of cell objects in the prediction mask. The second single-task network learned cell counting on the entire training set $D = D_1 \cup D_2$. In training, the annotated point counts were used as ground truths for images in $D_1$, and the cell counts obtained by eyeballing were used as ground truths for images in $D_2$. Tables~\ref{table:serous-table} and~\ref{table:monuseg-table} show that the multitask networks outperformed their single-task counterparts, especially in cell counting. This improvement can be attributed to the shared encoder, which helps learn more representative features when cell localization is used as a complementary task. 

\begin{table*}[t]
\centering
\caption{Test set/fold results obtained by the proposed \textit{MixedSupervision} approach and the comparison algorithms when $p$ percent training data had point annotations. These are averages of 9 runs from 3 folds for the serous carcinoma dataset, and averages of 3 runs for the MoNuSeg dataset (as there was a single test set for the MoNuSeg dataset).}
\begin{tabular}{|c|l|c|c|c|c|c|c|c|c|}
\hline
\multirow{2}{*}{p(\%)} & \multirow{2}{*}{Model} & \multicolumn{3}{c|}{Serous carcinoma dataset} & \multicolumn{3}{c|}{MoNuSeg dataset} \\ \cline{3-8}
&  & F1-Score & $\text{RD}_\text{Loc}$ & $\text{RD}_\text{Count}$ &
     F1-Score & $\text{RD}_\text{Loc}$ & $\text{RD}_\text{Count}$ \\ \hline
\multirow{3}{*}{100} & \textit{MixedSupervision} & 84.45 $\pm$ 1.84 & 0.14 $\pm$ 0.04 & 0.19 $\pm$ 0.05 & 80.86 $\pm$ 0.44 & 0.07 $\pm$ 0.01 & 0.14 $\pm$ 0.02 \\ 
& ConCORDe-Net & 82.29 $\pm$ 2.60 & 0.16 $\pm$ 0.03 & 0.26 $\pm$ 0.15 & 74.37 $\pm$ 0.70 & 0.11 $\pm$ 0.02 & 0.19 $\pm$ 0.02 \\
& SSRNet & 55.60 $\pm$ 2.04 & 0.29 $\pm$ 0.01 & 0.34 $\pm$ 0.04 & 50.98 $\pm$ 1.29 & 0.21 $\pm$ 0.01 & 0.27 $\pm$ 0.03 \\ \hline
\multirow{2}{*}{75} & \textit{MixedSupervision} & 83.87 $\pm$ 1.80 & 0.13 $\pm$ 0.03 & 0.20 $\pm$ 0.06 & 79.10 $\pm$ 0.55 & 0.07 $\pm$ 0.03 & 0.15 $\pm$ 0.03 \\ 
& ConCORDe-Net & 79.92 $\pm$ 2.15 & 0.18 $\pm$ 0.02 & 0.30 $\pm$ 0.02 & 72.55 $\pm$ 0.93 & 0.12 $\pm$ 0.02 & 0.20 $\pm$ 0.02 \\ \hline
\multirow{2}{*}{50} & \textit{MixedSupervision} & 82.77 $\pm$ 1.82 & 0.14 $\pm$ 0.02 & 0.22 $\pm$ 0.02 & 77.68 $\pm$ 0.28 & 0.09 $\pm$ 0.03 & 0.19 $\pm$ 0.03 \\ 
& ConCORDe-Net & 77.59 $\pm$ 1.63 & 0.19 $\pm$ 0.05 & 0.31 $\pm$ 0.03 & 70.94 $\pm$ 1.12 & 0.14 $\pm$ 0.05 & 0.23 $\pm$ 0.02 \\ \hline
\multirow{2}{*}{25} & \textit{MixedSupervision} & 81.45 $\pm$ 1.95 & 0.15 $\pm$ 0.03 & 0.28 $\pm$ 0.05 & 76.41 $\pm$ 0.11 & 0.09 $\pm$ 0.03 & 0.18 $\pm$ 0.01 \\ 
& ConCORDe-Net & 74.18 $\pm$ 1.75 & 0.27 $\pm$ 0.04 & 0.40 $\pm$ 0.12 & 66.17 $\pm$ 1.73 & 0.18 $\pm$ 0.02 & 0.38 $\pm$ 0.04 \\ \hline
%
%
\end{tabular}
\label{table:comparison table}
\end{table*}

\begin{figure*}[t!]
    \centering
    \includegraphics[width=14cm]{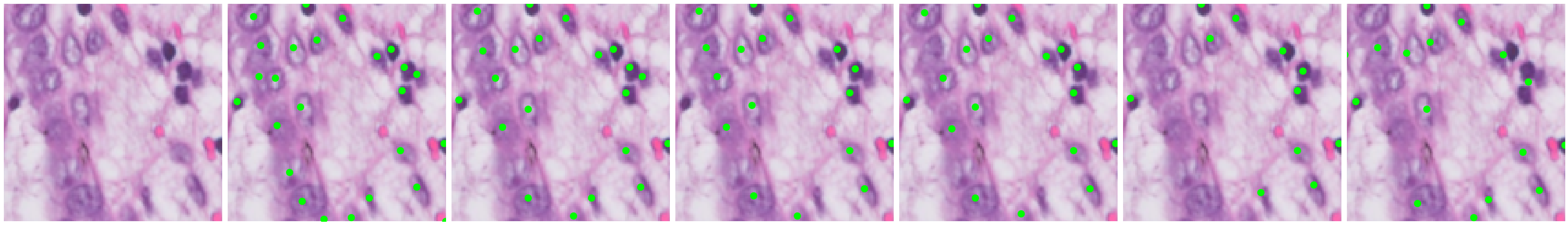} \\
    \includegraphics[width=14cm]{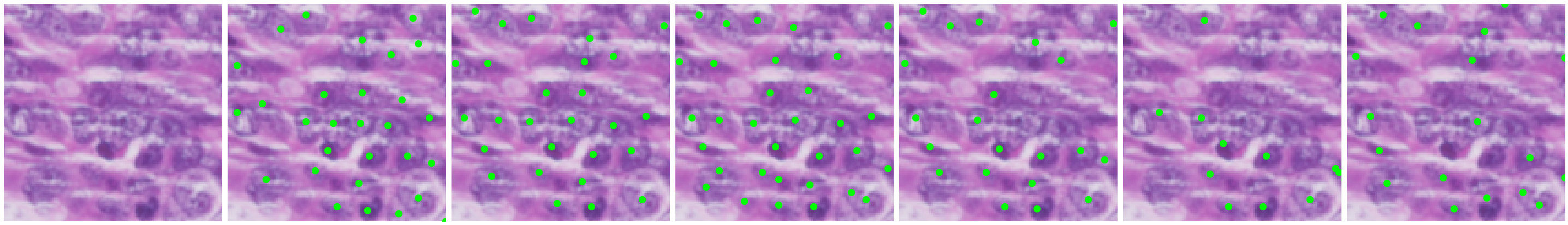} \\
    \includegraphics[width=14cm]{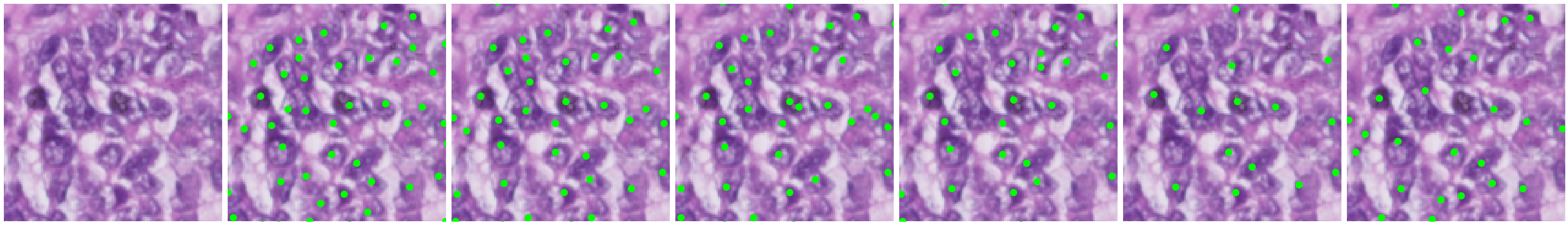} \\
    \includegraphics[width=14cm]{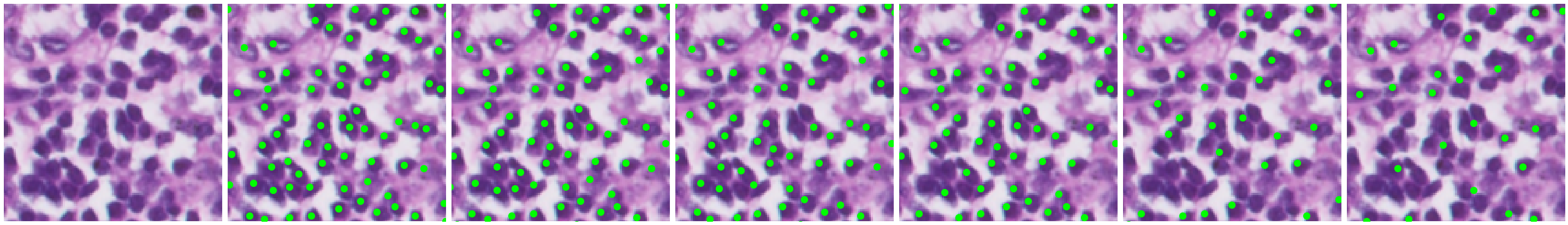} \\
    \includegraphics[width=14cm]{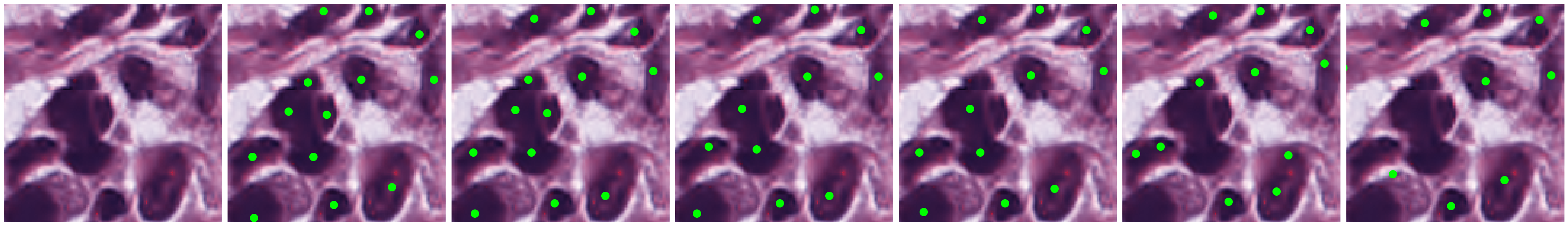} \\
    \includegraphics[width=14cm]{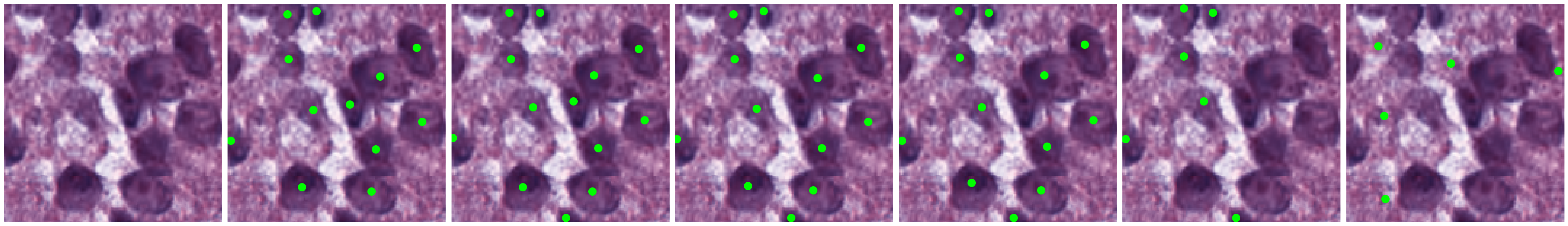} \\
    \includegraphics[width=14cm]{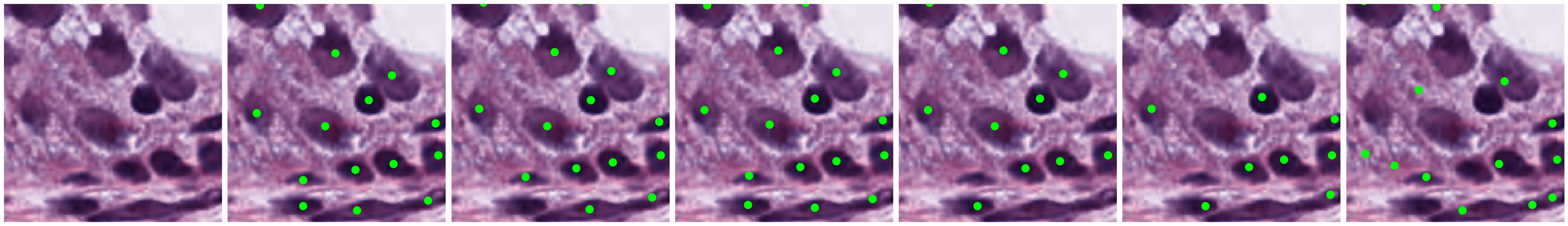} \\
    \includegraphics[width=14cm]{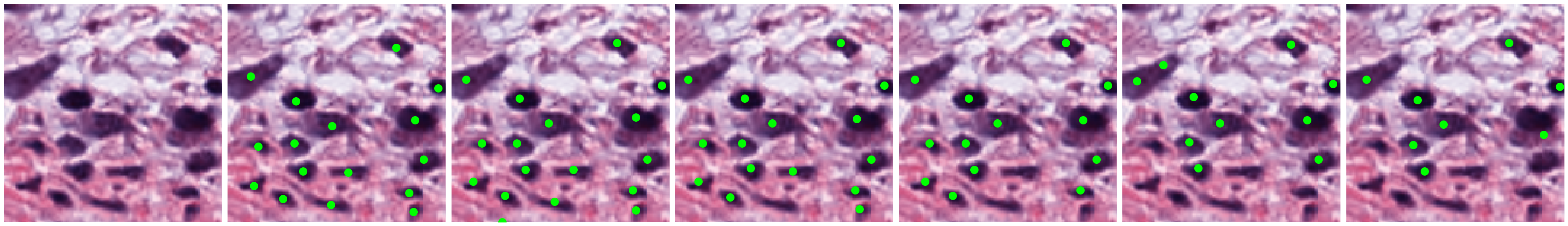} \\
    (a)\hspace{1.56cm}(b)\hspace{1.56cm}(c)\hspace{1.56cm}
    (d)\hspace{1.56cm}(e)\hspace{1.56cm}(f)\hspace{1.56cm}(g)\\
    \caption{Visual results on example test set images. (a) Patches cropped from the original images. The first four patches were selected from the serous carcinoma dataset, and the last four from the MoNuSeg dataset. The patches were cropped for better illustration. (b) Point annotations in the ground truths. (c) \textit{MixedSupervision} when $p=100$, (d) \textit{MixedSupervision} when $p=25$, (e) ConCORDe-Net~\cite{hagos2019concordenet} when $p=100$, (f) ConCORDe-Net when $p=25$, and (g) SSRNet~\cite{deng2023ssrnet}.}
    \label{fig:visual-results}
\end{figure*}

Next, we analyzed the effects of using the consistency term in the joint loss function. Tables~\ref{table:serous-table} and~\ref{table:monuseg-table} also report the results obtained without using the consistency term. They revealed that adding a consistency term to the combined loss function improved both cell localization and cell counting. This indicates the effectiveness of incorporating an additional consistency term in the cell counting branch to regularize the learning of the cell count prediction task, as well as to enhance the learning of the shared encoder when point annotations were unavailable in the training set. The results highlighted the benefits of using this loss function and the multitask design under data scarcity.

All these results suggested that good performance can be achieved even when only 25 percent of training images had point annotations, reducing the annotation effort by 75 percent. In the next subsection, we compared it with the \textit{ConCORDe-Net} algorithm, evaluating \textit{ConCORDe-Net} also for $p = \{100, 75, 50, 25\}$ percent. Since the \textit{SSR-Net} algorithm is a non-point annotation method, we trained it using the complete training set.

\subsection{Comparisons}

The quantitative results on the test sets are presented in Table~\ref{table:comparison table}. We first compared our proposed \textit{MixedSupervision} approach with \textit{ConCORDe-Net}, which uses a cascaded multi-task network with a different loss function for its cell counting task. Our approach achieved better object-level F1-scores and lower relative difference errors. The relative improvement was even higher as $p$ decreased. This might suggest that our proposed loss function, with the inclusion of the consistency term, facilitated the learning of a more robust and representative shared encoder, thereby enhancing the effectiveness of both tasks, even with the scarcity of strongly annotated data. Figure~\ref{fig:visual-results} presents visual results of this improvement. Note that these are smaller patches cropped from the original images for better illustration. As seen in this figure, although \textit{ConCORDe-Net} correctly identified most of the cells when all training data had point annotations (Fig.~\ref{fig:visual-results}e), the accuracy decreased when $p = 25$ percent (Fig.~\ref{fig:visual-results}f). On the other hand, the proposed \textit{MixedSupervision} approach led to better results when $p = 100$ percent, as well as the accuracy decrease was smaller when $p$ decreases to 25 percent (Figs.~\ref{fig:visual-results}c and~\ref{fig:visual-results}d). The second comparison model was \textit{SSRNet}, which also employs a multi-task architecture with a shared encoder. However, the results indicated that its integrated image reconstruction module, designed for non-point-based cell distribution, performed poorly on HE images, where the inter-cellular tissue serves as a diverse and complex background. This complexity caused reconstructed images to fail in providing accurate segmentation maps, leading to undersegmented cells (Fig.~\ref{fig:visual-results}g). 

\section{Conclusion}
This paper introduced a multitask network design employing a mixed-supervised training approach to simultaneously learn the tasks of cell counting and cell localization. The proposed \textit{MixedSupervision} approach relied on using two types of ground truth labels: point annotations to guide cell localization, and eyeballing-derived cell counts, which represented the weakest and least labor-intensive form of annotation, to supervise cell counting. To ensure consistency between the predictions of these two tasks, the \textit{MixedSupervision} approach introduced a consistency term in the definition of its loss function. We tested our proposed approach on two datasets of hematoxylin-eosin stained tissue images. Our experiments demonstrated that defining cell count prediction as an auxiliary task within a multitask network and using the eyeballing-derived cell counts as an additional supervisory signal in its training effectively regularized the cell localization task, even when stronger point annotations were limited. As a result, it led to better performance compared to its counterparts. 

This study is the first proposal of integrating eyeballing-derived ground truths into network training. It highlights the potential to increase the availability and the amount of supervisory signals for training networks in digital pathology, without relying on resource-intensive annotations. In future work, the potential of mixed-supervision approaches can be further explored and applied to a diverse range of tasks in digital pathology, unlocking new opportunities for innovation and advancement in the field. For instance, tumor diagnosis or classification, which "partially" rely on conventional eyeballing techniques, can be significantly improved by training networks with a mixed-supervision approach. This innovative strategy has the potential to revolutionize the field of digital pathology, enabling more accurate and efficient diagnoses/prognosis.

\bibliographystyle{IEEEtran} 
\bibliography{references}

\end{document}